\documentstyle[aps,eqsecnum]{revtex}
\begin{document}

\title{{\normalsize
\begin{flushright}
UB--ECM--PF--97/7\\
ULB--TH--97/9\\
hep-th/9705007\\
to appear in Nucl.\ Phys.\ B (Phys.\ Math.)\end{flushright}}
Extended Antifield--Formalism \footnote{
             This paper supersedes ``Ward Identities for Rigid Symmetries
             of Higher Order", hep-th/9611056} } 

\author{Friedemann Brandt $^a$ \footnote{E-mail: brandt@ecm.ub.es}, 
Marc Henneaux $^{b,c}$ \footnote{E-mail: henneaux@ulb.ac.be} and 
Andr\'{e} Wilch $^b$ \footnote{E-mail: awilch@ulb.ac.be} }

\address{$^a$ Departament d'Estructura i Constituents de la Mat\`eria,
Facultat de F\'{\i}sica,\\
Universitat de Barcelona,
Diagonal 647, E--08028 Barcelona, Spain.\\
$^b$ Facult\'e des Sciences, Universit\'e Libre de Bruxelles,\\
Campus Plaine C.P. 231, B--1050 Bruxelles, Belgium.\\
$^c$ Centro de Estudios Cient\'\i ficos de Santiago,\\
Casilla 16443, Santiago 9, Chile}

\maketitle

\begin{abstract}
The antifield formalism is extended so as to incorporate 
the rigid symmetries of a given theory.  To that
end, it is necessary to introduce global ghosts not only
for the given rigid symmetries, but also for all the
higher order conservation laws, associated with conserved 
antisymmetric tensors $j^{\mu_1 \ldots \mu_k}$ fulfilling
$\partial_{\mu_1} j^{\mu_1 \ldots \mu_k} \approx 0$.
Otherwise, one may encounter obstructions of the
type discussed in \cite{global}.
These higher order conservation laws are shown to define additional 
rigid symmetries of the master equation and to form -- together with
the standard symmetries -- an interesting algebraic structure. 
They lead furthermore to independent 
Ward identities which are derived in the standard manner, because 
the resulting master (``Zinn-Justin") equation capturing both the
gauge symmetries and the rigid symmetries of all orders takes
a known form. 
Issues such as anomalies or consistent
deformations of the action preserving some set of rigid symmetries
can be also systematically analysed in this framework. 
\end{abstract}

%%%%%%%%%%%%%%%%%%%%%%%%%%% body of paper %%%%%%%%%%%%%%%%%%%%%%%%%%%%%%%%

%=========================================================================
\section{Introduction}
%=========================================================================

Gauge theories may possess global (= rigid) symmetries
in addition to their local (= gauge) symmetries. Through Noether's first
theorem, non-trivial rigid symmetries of the classical action correspond
to non-trivial conserved currents,
\begin{equation}
\partial_\mu j^\mu \approx 0\;.
\label{cons1}
\end{equation}
Now, (\ref{cons1}) is actually only a special case of the more general 
conservation law
\begin{equation}
\partial_{\mu_1} j^{\mu_1  \dots \mu_k}  \approx 0
\label{cons2}
\end{equation}
where  $j^{\mu_1  \dots \mu_k}$ is  completely
antisymmetric and the symbol
$\approx$ denotes weak (i.e. on-shell) equality.  Non-trivial
solutions of (\ref{cons2}) define what we
call non-trivial conservation laws of order $k$.
\footnote{A conserved antisymmetric tensor is trivial if it is of the form
          $j^{\mu_1 \dots \mu_k} 
          \approx \partial_{\nu} k^{\nu \mu_1 \dots \mu_k}$
           where $k^{\nu \mu_1 \dots \mu_k}$ is also completely 
           antisymmetric.}

Although it has been proved under fairly general conditions
that non-trivial conservation laws of higher order $k>1$
are absent for theories without gauge invariance
\cite{Vinogradov,Bryant,bbh1},
they may be present in the case of gauge theories.
Examples are given by $p$-form gauge theories which admit
non-trivial conserved antisymmetric tensors of rank $p+1$ \cite{hks}.
These conservation laws play an important role in supergravity 
(see e.g.\ \cite{Page}).  

The quantum mechanical implications of the rigid symmetries that are 
associated with ordinary conserved currents (\ref{cons1}) 
are well understood.  If these symmetries are non-linear, they get 
renormalized.   The most expedient way to derive the corresponding Ward
identities is to introduce sources for the composite operators 
representing the variations of the fields \cite{ZJ}.  In order to avoid an
infinite number of such sources (one for the first
variation, one for the second variation etc. \cite{Bonneau}),
constant ghosts are introduced of ghost number $1$ and of 
Grassmann parity opposite to that of the symmetry parameter \cite{Blasi}. 
The Ward identities then follow by solving an extended master equation
\cite{Blasi,Piguet1,Piguet2},
the explicit form of which will be given below.  This approach,
which works even if the gauge-fixing procedure does not
preserve manifest invariance under the rigid symmetry, has
proved useful in the investigation of the renormalization
and anomaly problems in globally supersymmetric models \cite{many,glusy}.

However, it has been shown that the construction of
a local solution of the extended master equation of \cite{Blasi} 
may get obstructed -- already at the classical level --
in presence of higher order non-trivial conservation laws \cite{global}.
When this occurs, it is not possible to incorporate the standard rigid
symmetries along the lines of \cite{Blasi} and the procedure breaks down.

The main purpose of this paper
is to show that the obstructions can be avoided and that locality 
can be recovered if one extends
the approach of \cite{Blasi} by properly including in the formalism also
the higher order conservation laws. Together with the original symmetries,
they form a rich algebraic structure involving structure constants of 
increasing order,
which fulfill generalized Jacobi identities. We introduce global
ghosts for each independent rigid symmetry of any order
and we write down the corresponding 
form of the master equation (equation (4.1) below). 
We then prove the existence of a local solution of the master
equation, which is our main result.

Because the master equation incorporating all the gauge and
rigid symmetries takes a form that is very similar to the
standard one, one can derive, by the familiar procedure of
differentiating it with respect to the sources, the
Ward identities for the Green functions.  In particular, since
the rigid symmetries of higher order come with their own
ghosts, they lead to independent Ward identities.
Similarly, the analysis of
anomalies in both the rigid and gauge symmetries may still be
formulated as a cohomological problem, with a differential
extending the standard BRST operator. 

We finally describe the consistent
deformations of the action in this context.

%=========================================================================
\section{Higher Order Conservation Laws as Higher Order Symmetries}
%=========================================================================

One could of course attempt to regard the higher order conserved antisymmetric 
tensors $j^{\mu_1 \ldots \mu_k}$ as ordinary Noether currents parametrized
by further indices. There are many good reasons for not doing this. 
One of them is that this approach does not yield the appropriate notion of
``triviality" because it does not take properly into account all the
antisymmetry properties of $j^{\mu_1 \ldots \mu_k}$. Viewed as a higher
order conservation law, the equation 
$\partial_{\mu_1} j^{\mu_1 \ldots \mu_k} \approx 0$
is trivial if and only if 
$j^{\mu_1 \ldots \mu_k} \approx \partial_{\nu} k^{\nu \mu_1 \cdots \mu_k}$,
where $k^{\nu \mu_1 \ldots \mu_k}$ is also completely antisymmetric.
But if one views $\partial_{\mu_1} j^{\mu_1 \ldots \mu_k} \approx 0$
as a collection of ordinary conservation laws parametrized by further
indices $\mu_2, \ldots, \mu_k$, triviality holds under the weaker condition
$j^{\mu_1 \ldots \mu_k} \approx \partial_{\nu} S^{\nu \mu_1 \cdots \mu_k}$,
where $S^{\nu \mu_1 \cdots \mu_k}$ is only required to be antisymmetric in
its first two indices $\nu$ and $\mu_1$.

That these two notions are inequivalent is best illustrated in the case 
of the free $n$-dimensional Maxwell theory, for which 
$\partial_{\mu_1} F^{\mu_1 \mu_2} \approx 0$, 
$F^{\mu_1 \mu_2} = - F^{\mu_2 \mu_1}$. 
The relation $\partial_{\mu_1} F^{\mu_1 \mu_2} \approx 0$ is a non-trivial
conservation law of order $2$ because there is no completely antisymmetric
$k^{\nu \mu_1 \mu_2}$ which is local (i.e. polynomial in derivatives)
and satisfies 
$F^{\mu_1 \mu_2} \approx \partial_{\nu} k^{\nu \mu_1 \mu_2}$,
even if one allows $k^{\nu \mu_1 \mu_2}$ to depend explicitly on $x^{\mu}$
\cite{bbh1}. In contrast to that, the $F^{\mu_1 \mu_2}$ become trivial when
they are regarded as a set of $n$ Noether currents, 
one for each value of $\mu_2$,
\begin{equation}
F^{\mu_1 \mu_2} \approx \partial_{\nu} S^{\nu \mu_1 \mu_2} \;, \ \
S^{\nu \mu_1 \mu_2} = x^{\mu_2} F^{\mu_1 \nu} = - S^{\mu_1 \nu \mu_2} \;.
\end{equation}
Accordingly, the conservation law 
$\partial_{\mu_1} F^{\mu_1 \mu_2} \approx 0$ does not correspond to
a nontrivial rigid symmetry of the Maxwell action. Rather,
it is associated with the shift 
symmetry $A_\mu\rightarrow A_\mu+\epsilon_\mu$ ($\epsilon_\mu=constant$)
which is a trivial symmetry as it is just a special gauge 
transformation of $A_\mu$
with parameter $x^\mu\epsilon_\mu$.

An extension of the Noether theorem that does take into account complete
antisymmetry has been proposed in the interesting work \cite{julia}.
We shall not follow that approach here, but rather, we shall directly 
relate the higher order conservation laws to 
{\em rigid symmetries of the solution of the master equation}. This point
of view turns out to be particularly convenient for the quantum theory.

Our starting point is thus the solution $S=S[\Phi^a,\Phi^*_a]$
of the master equation for
the gauge symmetries \cite{bv},
\begin{equation}
(S,S) = 0,
\label{master1}
\end{equation}
where $(\ ,\ )$ is the standard antibracket.
The $\{\Phi^a\}$ are the fields
(classical fields $\phi^i$, ghosts for the
gauge symmetries $C^{\alpha}$, ghosts of ghosts if
necessary, antighosts, Nakanishi-Lautrup auxiliary fields) while the
$\{\Phi^*_a\}$ are the corresponding antifields. The
master equation (\ref{master1}) always admits a solution
$S$ which is a {\em local} functional \cite{hen91}.

As shown in \cite{bbh1}, one can associate with each 
conservation law 
$\partial_{\mu_1} j_A^{\mu_1  \dots \mu_{k_A}}  \approx 0$ 
of order $k_A$ a local functional $S_A[\Phi^a, \Phi_a^*]$ 
which (i) has ghost number $-k_A$; and
(ii) is BRST-invariant. 
Since the BRST transformation $s$ is generated
in the antibracket by $S$, (ii) means 
\begin{equation}
s\,  S_A \equiv (S_A,S) =0.
\label{inv2}
\end{equation}
But this condition expresses at the same time 
that the solution $S$ of the master equation
is invariant under the canonical transformation generated in
field-antifield space by $S_A$,
\begin{equation}
\delta_A S  \equiv (S,S_A) =0.
\label{inv1}
\end{equation}
Consequently, {\em each conservation law defines indeed 
a symmetry of $S$}.

The relationship
between the conserved antisymmetric tensors and the generators 
$S_A$  has been given in \cite{bbh1}:
since $S_A=\int \mbox{d} x \, m_A$ is BRST invariant%
\footnote{Throughout the paper we use $\int \mbox{d} x$
to indicate integration over an $n$-dimensional 
base manifold (``spacetime''), and call a functional
BRST invariant if the BRST variation of its integrand
is a total derivative.}, its
integrand satisfies
\begin{equation}
sm_A + \partial_\mu m_A^\mu = 0.
\label{cons3}
\end{equation}
If $S_A$ has ghost number $-1$ (corresponding to ordinary rigid symmetries),
$m_A^\mu$ has ghost number zero and the antifield independent part of
(\ref{cons3}) reproduces (\ref{cons1}) because the antifield
independent part of $sm_A$ vanishes on-shell.
For $k=2$, the relation (\ref{cons3}) also holds,
but now $m_A$ has ghost number $-2$ and
$m_A^\mu$  has ghost number $-1$.  To get the conservation law
in the form (\ref{cons2}), the descent equation technique has to be used,
i.e. the differential $s$ has to be applied to (\ref{cons3}).  
Following standard arguments, this yields 
\begin{equation}
sm_A^\mu + \partial_\nu m_A^{\mu \nu} =0
\label{cons4}
\end{equation}
for some antisymmetric tensor $m_A^{\mu \nu}$ of ghost number
zero.  The conservation law (\ref{cons2}) for $k=2$ is just the
antifield-independent part of (\ref{cons4}), since
the antifield-independent part of $sm_A^\mu$ vanishes on-shell.
Similar arguments hold for the subsequent conservation laws with $k>2$.
By using the
antifield-BRST formalism, one can consequently provide a unified treatment
for all conservation laws.
(How to deal with descent equations
in the quantum theory is discussed in \cite{Lucchesi,Piguet1}).

%=========================================================================
\section{Algebraic Structure}\label{structure}
%=========================================================================

The symmetry generators $S_A$ possess an interesting algebraic
structure.  Assume that $\{S_A\}$ is a basis of symmetry generators, i.e.
of the cohomology $H^{-k}(s)$ in the
space of local functionals $\Gamma[\Phi,\Phi^*]$ at all negative
ghost numbers $(-k)$.
This means that any $s$-closed local functional $\Gamma[\Phi,\Phi^*]$
with negative ghost number is a linear combination of the
$S_A$ up to an $s$-exact term,
\begin{equation} 
s \Gamma[\Phi,\Phi^*]=0 \ \mbox{and} \
gh(\Gamma) < 0 \ \Leftrightarrow \
\Gamma[\Phi,\Phi^*]=\lambda^A S_A[\Phi,\Phi^*] +
s \tilde\Gamma[\Phi,\Phi^*] \;,
\label{4a}
\end{equation}
and no non-vanishing linear combination of the $S_A$ is
$s$-exact in the space of local functionals
$\Gamma[\Phi,\Phi^*]$,
\begin{equation} 
\lambda^A \,S_A[\Phi,\Phi^*] = s \Gamma[\Phi,\Phi^*]
\ \Leftrightarrow \ \lambda^A = 0 \quad \forall A\;.
\label{4b}
\end{equation}
Since the antibracket of two $S_A$ is BRST-closed (one
has $(S,(S_A,S_B))=0$ by
the Jacobi identity for the antibracket),
it must be of the form
\begin{equation}
(-1)^{\varepsilon_A}(S_A,S_B) = f_{AB}^D S_D + (S,S_{AB}) 
\label{16a}
\end{equation}
for some constants $f_{AB}^C$ and some local functionals
$S_{AB}$ ($\varepsilon_A+1$ denotes the Grassmann parity of
$S_A$; phases and factors are introduced for later convenience).  
Taking the antibracket of this expression with $S_C$
and using the Jacobi identity for the antibracket then leads to
\begin{equation}
S_Ef_{D[A}^Ef_{BC]}^D=
\left(S,(-)^{\varepsilon_B}(S_{[B},S_{CA]})-S_{D[A}f_{BC]}^D\right)
\label{basis4}
\end{equation}
where $[\;]$ denotes graded antisymmetrization.
According to (\ref{4b}), both sides of (\ref{basis4}) have to 
vanish separately.
This yields the Jacobi identity for the structure constants $f^C_{AB}$ 
and -- due to (\ref{4a}) --
the additional identity
\begin{equation}
(-)^{\varepsilon_A}(S_{[A},S_{BC]})=S_{D[C}f_{AB]}^D+
\mbox{\small{$\frac{1}{3}$}}f_{ABC}^DS_D
+\mbox{\small{$\frac{1}{3}$}}(S,S_{ABC})
\label{algebra1}
\end{equation}
for some second order structure constants $f_{ABC}^D$ and
some local functionals $S_{ABC}$.
If there does not exist any higher order conservation law (and thus no
$S_A$ with $gh(S_A)<-1$), then higher order structure constants like
$f_{ABC}^D$ cannot occur.
This follows from a mere ghost number counting argument.  
However, in the presence of higher order symmetries, terms of the
form $f_{ABC}^DS_D$ are allowed and indeed do occur in explicit
examples (see \cite{global} and the example treated below).

The above construction can be continued, defining further local functionals
$S_{A_1 \cdots A_r}$ and structure constants $f^C_{A_1 \cdots A_r}$
that will satisfy generalized higher order Jacobi identities.
As an illustration, we just provide the next step, 
leading to the local functionals
$S_{ABCD}$ and the structure constants $f^E_{ABCD}$:
By taking the antibracket of (\ref{algebra1}) with $S_D$ and using the 
Jacobi-identity, one obtains
\begin{eqnarray}
\lefteqn{ \mbox{\small{$\frac{1}{12}$}}(S,S_{ABCD}) + 
          \mbox{\small{$\frac{1}{3}$}}
          (S_{[A},S_{BCD]})(-)^{\varepsilon_A + 1} +
          \mbox{\small{$\frac{1}{4}$}}
          (S_{[AB},S_{CD]})(-)^{\varepsilon_A + \varepsilon_B} }\nonumber\\
& & + \mbox{\small{$\frac{1}{12}$}}f_{ABCD}^E S_E +
      \mbox{\small{$\frac{1}{3}$}}S_{E[D} f_{ABC]}^E +
      \mbox{\small{$\frac{1}{2}$}}S_{E[CD} f_{AB]}^E = 0
\label{algebra2}      
\end{eqnarray}
and
\begin{equation}
\mbox{\small{$\frac{1}{2}$}} f^F_{E[CD} f^E_{AB]} +
\mbox{\small{$\frac{1}{3}$}} f^F_{E[D} f^E_{ABC]} = 0 \;.
\label{generaljacobi1}
\end{equation}
Eq.(\ref{generaljacobi1}) is a generalized Jacobi identity
for the higher order structure constants $f^E_{ABC}$. 
Both the usual Jacobi identity and Eq.(\ref{generaljacobi1}) can be written as
\begin{equation} 
\sum_{r=2}^{p-1}
\frac{1}{r!\,(p-r)!}\, 
%\mbox{\small{$\frac{1}{r!(p-r)!}$}}
f_{C[A_{r+1} \cdots A_p}^D f_{A_1\cdots A_r]}^C = 0
\label{generaljacobi2}
\end{equation}
where $p=$3 and 4. It turns out that the Jacobi identities for the subsequent
structure constants are also given by (\ref{generaljacobi2}).
Note that (\ref{16a}) contains the commutation relations
of the standard rigid symmetries (for $k_A=k_B=1$), and
that (\ref{generaljacobi2}) includes the
Jacobi identities for the corresponding structure constants.

Below, we shall set up the extended antifield formalism 
such that it automatically incorporates this algebra 
to all orders through a modified master equation.
Algebraic structures similar to the ones appearing here 
have been analyzed in \cite{azc}. 

A subset of symmetry generators $S_\alpha$ defines a 
subalgebra if and only if the relations to which they lead never involve
the other symmetry generators $S_\Delta$, $A=(\alpha, \Delta)$. 
An equivalent condition is that
the structure constants $f_{\alpha_1 \alpha_2}^\Delta$,
$f_{\alpha_1 \alpha_2 \alpha_3}^\Delta$ $\dots$ all vanish.
The subset
$\{S_\alpha$, $S_{\alpha_1 \alpha_2}$, $S_{\alpha_1 \alpha_2 \alpha_3}$,
$\dots\}$ is then a closed set for the generating equations
(\ref{16a}) and the subsequent ones.  As
shown in \cite{global} and in the example below, 
the set of all symmetry generators of
order one (standard rigid symmetries) may not form a subalgebra in
the above sense.

%=========================================================================
\section{Extended Master Equation}
%=========================================================================

The most expedient way of generating all the local functionals 
$S_{A_1 \cdots A_r}$ and structure functions $f^C_{A_1 \cdots A_r}$
appearing in the algebra described above, is through an extended master
equation. Another motivation for using the 
master-equation approach has to do with the quantum theory.
The fact that all the conservation laws, including the
higher-order ones, appear as symmetries of the solution
of the usual master equation, makes it possible to investigate in
a unified manner the corresponding Ward identities.  Since
the transformations generated by the $S_A$ may be non-linear,
they may get renormalized in the quantum theory.  To cope
with this feature, 
we extend the  approach of \cite{ZJ,BRS,Blasi} and introduce,
besides the standard antifields and local ghosts associated with
the gauge symmetry,
constant ghosts $\xi^A$ for {\em all} independent local conservation laws.
The constant ghosts are
assigned opposite ghost number and 
the same Grassmann parity as the corresponding generator
$S_A$. Consequently, the ghost number of $\xi^A$ equals the
order $k_A$ of the corresponding conservation law,
\[
gh(\xi^A) = - gh(S_A) = k_A \;.
\]
For instance, constant ghosts
corresponding to Noether currents carry ghost number 1, as
one expects since these ghosts correspond
to global symmetries of the classical action.
\footnote{ It is crucial, in order to avoid the
           obstructions, to take the higher order conservation 
           laws into account as done here, with constant ghosts of
           ghost number $k$. It would not work to 
           treat the higher order conservations laws (\ref{cons2})
           as ordinary conservation laws parametrized by further
           indices and to associate with them constant ghosts
           $\xi^{\mu_1 \dots\mu_k}$ of ghost number one.}
  
We then add the
term $S_A \xi^A$ to $S$ and search for a solution ${\cal S}[\Phi,
\Phi^*, \xi]$ of
the extended master equation
\begin{equation}
({\cal S},{\cal S}) + 2\sum_{r\geq 2}\frac 1{r!}\,
\frac{\partial^R {\cal S}}{\partial \xi^B} \, f_{A_1\cdots A_r}^{B}
\xi^{A_r}\cdots\xi^{A_1} = 0
\label{extendedmaster1}
\end{equation}
of the form
\begin{equation}
{\cal S} = S + S_A \xi^A + \sum_{r\geq 2}
\frac 1{r!}\, S_{A_1\cdots A_r}\xi^{A_r}\cdots\xi^{A_1},
\label{formofS}
\end{equation}
where the $f_{A_1\cdots A_r}^{B}$ are
the structure constants and the $S_{A_1\cdots A_r}$ are the local
functionals of the symmetry algebra described above, and still need to be
constructed.

The existence-proof of ${\cal S}$, to be given in the
next section, becomes straightforward
if constant antifields $\xi_A^*$ conjugate to the 
constant ghosts $\xi^A$
are introduced through
\begin{equation} {\cal S}'=
{\cal S} + \sum_{r\geq 2}\frac 1{r!}\,\xi^*_B\, f_{A_1\cdots A_r}^{B}
\xi^{A_r}\cdots\xi^{A_1} \;.
\label{formofS'}
\end{equation}
These additional antifields have the usual properties,
\[
gh(\xi_A^*) = - gh(\xi_A) - 1 = - k_A - 1
\]
and
\[
\varepsilon(\xi^*_A) = \varepsilon(\xi_A) + 1 = \varepsilon_A \;.
\] 
The extended master equation 
(\ref{extendedmaster1}) now takes
the familiar form  
\begin{equation}
({\cal S}', {\cal S}')' = 0 \;,
\label{extendedmaster2}
\end{equation}
where the extended antibracket $(\ ,\ )'$ is given by
\[
 (X,Y)'=
 \frac{\partial^R X}{\partial\xi^A}
 \frac{\partial^L Y}{\partial\xi^*_A}-
 \frac{\partial^R X}{\partial\xi^*_A}
 \frac{\partial^L Y}{\partial\xi^A}
 +\int \mbox{d} x \left[
 \frac{\delta^R X}{\delta\Phi^a(x)}
 \frac{\delta^L Y}{\delta\Phi^*_a(x)}-
 \frac{\delta^R X}{\delta\Phi^*_a(x)}
 \frac{\delta^L Y}{\delta\Phi^a(x)}\right].
\]
%====================================
\section{Existence of ${\cal S}$}
%====================================

In order to prove that there always exists a solution of 
Eq.(\ref{extendedmaster2}), we shall follow the method of \cite{HPT2}.
For this purpose we shall extend the definition of
the Koszul-Tate differential $\delta$ appropriately
and use an expansion of
(\ref{extendedmaster2}) according to the antighost number ($agh$)
defined by
\[ agh(\Phi^a)=agh(\xi^A)=0,\quad agh (\Phi^*_a)=-gh(\Phi^*_a),
\quad agh(\xi^*_A)=-gh(\xi^*_A)=k_A+1.
\]
Before we do this, we recall some standard results
\cite{bbh1,hen91,HPT2} on the 
cohomology of the Koszul-Tate operator in the space ${\cal F}$
of local functionals $\Gamma[\Phi,\Phi^*]$.
On the fields $\Phi^a$ and their antifields $\Phi^*_a$, 
$\delta$ is defined through \cite{HPT2}
\begin{eqnarray} 
\delta\, \Phi^a=0,\quad 
\delta\, \phi^*_i=-\frac{\delta^L S_0}{\delta\phi^i}\ ,
\quad\cdots
\label{KT}\end{eqnarray}
where $S_0$ is the classical action. 
Now, in ${\cal F}$, the cohomologies of $s$ and $\delta$ are
isomorphic at all negative ghost numbers $(-k)$ and
positive antighost numbers $k$ respectively,
$H^{-k}(s,{\cal F})\simeq H_{k}(\delta,{\cal F})$ for $k>0$
(superscript and subscript of $H$ denote the ghost number and 
antighost number respectively).
The representatives $S_A^0$ of
$H_{k}(\delta,{\cal F})$, $k>0$ can be chosen so as not to depend
on the ghost fields. The corresponding
representatives $S_A$ of $H^{-k}(s,{\cal F})$
are BRST-invariant extensions of the $S_A^0$,
\begin{equation} 
S_A[\Phi, \Phi^*] = S_A^0[\phi, \Phi^*] + \mbox{ghost-terms}.
\label{completion}\end{equation} 
Note that the part of $S_A$ that does not involve the ghosts satisfies 
\[ 
gh(S_A^0)=-agh(S_A^0)=-k_A \;. 
\]
The $S_A^0$ fulfill therefore requirements analogous to
(\ref{4a}) and (\ref{4b}), i.e.
\begin{eqnarray} 
& &\delta \Gamma[\Phi,\Phi^*]=0 \ \mbox{and} \
agh(\Gamma) > 0 \ \Leftrightarrow \
\Gamma[\Phi,\Phi^*]=\lambda^A S_A^0[\phi,\Phi^*] +
\delta \tilde\Gamma[\Phi,\Phi^*] \;,\mbox{\ }
\label{5a}\\
& &\lambda^A \,S_A^0[\phi,\Phi^*] = \delta \Gamma[\Phi,\Phi^*]
\ \Leftrightarrow \ \lambda^A = 0 \quad \forall A\;.
\label{5b}
\end{eqnarray}

Now we define the above-mentioned extension of $\delta$.
It applies to 
the functional space to which ${\cal S}'$ belongs, namely the 
vector space ${\cal E}$ of functionals ${\cal A}$ defined through
\begin{equation}
{\cal A}\in{\cal E}\quad:\Leftrightarrow\quad
{\cal A}=\Gamma[\Phi,\Phi^*,\xi] +\lambda^A(\xi)\, \xi^*_A \;,
\end{equation}
where $\Gamma=\int \omega_n$ is an integrated local
volume form which does not
involve the $\xi^*_A$ (it may depend polynomially on the $\xi^A$) and
$\lambda^A(\xi)$ is a polynomial in the constant ghosts. Note that
functionals in ${\cal E}$ depend on the $\xi^*_A$ at most
{\em linearly} via non-integrated terms $\lambda^A(\xi)\, \xi^*_A$.

To define $\delta$ in ${\cal E}$ appropriately, we
extend its definition to the global ghosts and antifields via
\begin{eqnarray}
\delta \, \xi^A =0,\quad
\delta \, \xi^*_A =  S_A^0[\phi, \Phi^*]
\label{5}
\end{eqnarray}
where $S_A^0$ is the ghost independent part of $S_A$,
see above. On the $\Phi^a$ and $\Phi^*_a$, $\delta$ is
defined as before in (\ref{KT}). Note that
$\delta$ is well-defined in ${\cal E}$, as
${\cal A}\in{\cal E}\Rightarrow \delta{\cal A}\in{\cal E}$.

With these definitions, $\delta$ is nilpotent due to
$ \delta^2\xi^*_A=\delta S_A^0=0$. Furthermore,
by construction, it is acyclic in ${\cal E}$
at positive antighost number, i.e.\ $H_k(\delta,{\cal E})=0$
for $k>0$.
Indeed, $\delta {\cal A}=0$ is equivalent to
$\delta \Gamma + \lambda^A S_A^0=0$ which implies $\lambda^A=0$
for $agh({\cal A})>0$ due to (\ref{5b}) and thus
also $\delta\Gamma=0$. From this we conclude, using (\ref{5a})
and (\ref{5}),
${\cal A}=\Gamma=\delta\tilde\Gamma + \tilde\lambda^A S_A^0
=\delta(\tilde\Gamma+\tilde\lambda^A\xi^*_A)$ and thus
\begin{eqnarray}
& &\delta {\cal A}=0,\quad agh( {\cal A})>0,\quad
{\cal A}\in {\cal E}\nonumber\\
& &\Rightarrow\quad
{\cal A}=\delta  \tilde{\cal A},\quad  \tilde{\cal A}\in {\cal E}.
\label{8a}
\end{eqnarray}

The construction of solutions to Eq.(\ref{extendedmaster2})
now follows almost word for word the standard pattern of homological
perturbation theory \cite{HPT2} (section 10.5.4).
The sought ${\cal S}'$ is expanded
according to the antighost number,
\begin{equation}
 {\cal S}' = S_0 + S_1 + S_2 + \cdots \;,\ agh(S_k)=k,\
 S_k\in {\cal E}.
\label{11}
\end{equation}
Here $S_0$ is the classical action and we require ${\cal S}'$ to
contain the piece $S_A^0 \xi^A$, i.e.
\begin{equation}
\frac{\partial^R{\cal S}'}{\partial\xi^A }= S_A^0 + O(k_A+1)
\label{11a}
\end{equation}
where $O(k)$ denotes collectively terms with antighost numbers $\geq k$.
Together with the standard conditions (``properness")
on the solution $S$ of the
usual master equation in gauge theories,
(\ref{11a}) fixes the boundary conditions that we impose on
${\cal S}$ in order to guarantee that it encodes indeed
all the local conservation laws. This fixes in particular $S_1$
to the form
\begin{equation} 
S_1= \phi^*_i\;R^i_{\alpha}\;C^{\alpha} +\sum_{A:k_A=1}S_A^0\xi^A
\end{equation}
where the first term encodes the gauge symmetries of $S_0$
(we used De Witt's notation)
and the second term contains its global symmetries (for $k_A=1$
one has $S_A^0 = \phi^*_i(\delta_A\phi^i)$ in the notation of
\cite{global} where $\delta_A$ are the global symmetries).

The invariance of $S_0$ under the gauge and global symmetries
encoded in $S_1$ then ensures that $S_0 + S_1$ fulfills
the extended master-equation
up to terms of antighost number $\geq 1$.
Suppose now that $S^k = S_0 + S_1 + \cdots + S_k\in{\cal E}$
had been constructed
so as to satisfy (\ref{extendedmaster2}) and (\ref{11a})
up to terms of antighost number $\geq k$,
\begin{eqnarray}
\left( S^k, S^k \right)' = R_k + O(k+1),\ agh(R_k)=k \;, 
\label{12}\\
\frac{\partial^R S^k}{\partial\xi^A} = S_A^0 + O(k_A+1) \quad
\forall A:k_A\leq k.
\label{144}
\end{eqnarray}
Taking the extended antibracket of (\ref{12}) with $S^k$ and
using the Jacobi identity for the extended 
antibracket as well as (\ref{144}),
it is possible to infer that $R_k$ is $\delta$-closed, $\delta R_k=0$.
Furthermore we have $R_k\in{\cal E}$
since the vector space ${\cal E}$ is invariant under
the extended antibracket
(${\cal A},{\cal B}\in{\cal E}$ $\Rightarrow$
$({\cal A},{\cal B})'\in{\cal E}$).
Due to $k>0$, (\ref{8a}) therefore guarantees that
$R_k$ is $\delta$-exact in ${\cal E}$,
\begin{equation}
R_k=-2\delta S_{k+1},\quad S_{k+1}\in{\cal E},
\label{12a}\end{equation}
for some $S_{k+1}$.
This in turn implies that $S^{k+1}=S_0+\ldots+S_{k+1}$
satisfies the extended master equation up to
terms of antighost number $\geq (k+1)$,
\begin{equation}
 \left( S^{k+1}, S^{k+1} \right)' = R_{k+1} + O(k+2)
\end{equation}
where $agh(R_{k+1})=k+1$.
Note that (\ref{12a}) determines
$S_{k+1}$ only up to a $\delta$-closed
functional in ${\cal E}$. In particular one can always add to
$S_{k+1}$ a term of the form $\sum_{A:k_A=k+1}S_A^0\xi^A$ without
violating (\ref{12a}). Hence, the ``boundary conditions''
(\ref{11a}) can always be fulfilled.
Since the arguments apply to all $k>0$, we have indeed
proved the existence of a solution ${\cal S}$
to the extended master equation
of the form (\ref{formofS}, \ref{formofS'}) with the required properties.
In other words, the inclusion of global ghosts for the rigid symmetries
of higher order eliminates the obstructions found in \cite{global}.

For the sake of completeness we remark that the boundary conditions
(\ref{11a}) guarantee that the part of ${\cal S}$ which
is linear in $\xi^A$ is indeed of the form $S_A\xi^A$ where
$S_A$ is a BRST-invariant completion of $S_A^0$, cf. (\ref{completion}).
Indeed, one has
\begin{equation} 
\left({\cal S}',{\cal S}'\right)'=0
\ \Rightarrow\
(\partial^R{\cal S}'/\partial \xi^A,{\cal S}')'=0
\ \Rightarrow\ 
(S_A,S)=0
\label{zusatz1}\end{equation} 
where 
\begin{equation}
S_A\equiv 
\frac{\partial {\cal S'}}{\partial \xi^A}\left|_{\xi=0}\right.
=\frac{\partial {\cal S}}{\partial \xi^A}\left|_{\xi=0}\right. .
\label{zusatz2}\end{equation}
In (\ref{zusatz1}) we first differentiated the extended master
equation with respect to $\xi^A$ and set all the $\xi^A$ to zero
afterwards. (\ref{inv2}),
(\ref{zusatz1}), (\ref{zusatz2}) and (\ref{11a}) show that 
the part $S_A$ of ${\cal S}$ which is linear in $\xi^A$ is 
indeed a BRST-invariant completion of $S_A^0$, as promised.

A remarkable feature of the extended master equation (\ref{extendedmaster1})
or (\ref{extendedmaster2})
is that it encodes the structure constants of the algebra
of rigid symmetries described above.  Indeed, given $S$ and the
generators $S_A\xi^A$ (or actually, just their pieces $S_A^0\xi^A$ linear
in the ghosts), the higher order functionals $S_{A_1 \dots A_r}$
and the structure constants $f^B_{A_1\dots A_r}$ are recursively
determined by the demand that the extended master 
equation be satisfied. For instance, one gets relation (\ref{algebra1})
by collecting in Eq.(\ref{extendedmaster2}) all the terms that contain
three global ghosts and no global antifield. The relation (\ref{algebra2})
then corresponds to the part containing four global ghosts and no $\xi^*$.
The terms involving $\xi^*$ provide the Jacobi identities 
(\ref{generaljacobi2}).

The fact that the extended master equation captures the complete algebraic 
structure of the gauge and the rigid symmetries parallels the property of 
the usual master equation that encodes all the information on
the algebra of gauge transformations, including the Jacobi identities
of first and higher order \cite{bv,HPT2}.
When there are only standard rigid symmetries, Eq.(\ref{extendedmaster1}) 
reduces to the extended master
equation of \cite{Blasi},
\begin{equation}
({\cal S},{\cal S}) + 
\frac{\partial^R {\cal S}}{\partial \xi^C} \, f_{BA}^{C}
\xi^{A}\xi^{B} = 0 \;.
\label{extended'}
\end{equation}
In the absence of rigid symmetries of any order, the extended master equation
reduces, of course, to (\ref{master1}).

%=========================================================================
\section{Ward Identities}
%=========================================================================

{}From the extended master equation (\ref{extendedmaster1}),
the Ward identitites for the Green functions can be derived straightforwardly.
Since the extended master equation (\ref{extendedmaster1})
is similar to the extended
master equation  (\ref{extended'}) of  \cite{Blasi}, the procedure 
follows the familiar pattern and we sketch only the main steps.

The generating functional for the Green functions of the
theory is given by the path integral
\begin{equation}
Z_{J,K,\xi} = \int [D\Phi] \exp i \{  {\cal S}^{\Psi}[\Phi, K, \xi] + 
\int \mbox{d} x J_a(x) \Phi^a(x) \}.
\end{equation}
The functional $ {\cal S}^{\Psi}[\Phi, K, \xi]$ appearing in $Z_{J,K,\xi}$ is
obtained from ${\cal S}[\Phi, \Phi^*, \xi]$
by making the transformation
$\Phi^*_a = K_a + \frac{\delta \Psi}{\delta \Phi^a}$, where the
gauge-fixing fermion $\Psi[\Phi]$ is chosen such that
${\cal S}^{\Psi}[\Phi, 0, 0]$ is completely gauge-fixed. 
The functional ${\cal S}^{\Psi}[\Phi, K, \xi]$ obeys
the same equation (\ref{extendedmaster1}) -- with $\Phi^*$ replaced by $K$ --
as ${\cal S}[\Phi,\Phi^*,\xi]$, because
the transformation from $\Phi^*$ to $K$ is a canonical transformation
that does not involve the $\xi^A$.  The fields
$J_a(x)$ and $K_a(x)$, as well as the constant ghosts $\xi^A$,
are external sources not to be integrated over in the path integral.
Now, perform in $Z_{J,K,\xi}$ the infinitesimal change of integration
variables
\begin{equation}
\Phi^a \rightarrow \Phi^a + ( \Phi^a,{\cal S}^{\Psi}) =
\Phi^a + \frac{\delta^L {\cal S}^{\Psi}}
{\delta K_a}.
\end{equation}
Using (\ref{extendedmaster1}) and assuming the measure
to be invariant
\footnote{If the measure is not invariant, one must replace
          the extended master equation (\ref{extendedmaster1}) by the
          quantum extended master equation, which reads
          \[
          ({\cal S},{\cal S}) + 2 i \Delta {\cal S} + 
          2\sum_{r\geq 2}\frac {1}{r!} \,
          \frac{\partial^R {\cal S}}{\partial \xi^B} \, f_{A_1\cdots A_r}^{B}
          \xi^{A_r}\cdots\xi^{A_1} = 0
          \]
          where of course a meaningful (regularized) definition has to be
          given to $\Delta$.  The Ward identitites
          (\ref{WI1}) and (\ref{WI3}) are unchanged in absence
          of anomalies.},
the following Ward identity results
for $Z_{J,K,\xi}$:
\begin{equation}
\int \mbox{d} x\, J_a(x)\, \frac{\delta^L Z_{J,K,\xi}}{\delta K_a(x)}
 \; - \; \sum_{r\geq 2}\frac {1}{r!} \,
\frac{\partial^R Z_{J,K,\xi}}{\partial \xi^B} \, f_{A_1\cdots A_r}^{B}
\xi^{A_r}\cdots\xi^{A_1}  = 0 \;.
\label{WI1}
\end{equation}
Since Eq.(\ref{WI1}) is a linear functional
equation on $Z_{J,K,\xi}$, the generating functional
$W = -i \ln Z_{J,K,\xi}$ for the connected Green functions
obeys the same identity.
Performing the standard Legendre transformation
\begin{eqnarray}
& &\Gamma[\Phi_c, K, \xi] = W[J,K,\xi] - \int \mbox{d} x J_a(x) \Phi^a_c(x)\ ,
\nonumber \\
& &\Phi^a_c(x) = \frac{\delta^L W}{\delta J_a(x)}\, , \; \;
J_a(x) = - \frac{\delta^R \Gamma}{\delta \Phi^a_c(x)}
\label{Gamma}
\end{eqnarray}
one finds that the effective action $\Gamma$
fulfills a Ward identity of the same form as (\ref{extendedmaster1}),
\begin{equation}
\int \mbox{d} x\, 
 \frac{\delta^R \Gamma}{\delta\Phi^a_c(x)}\,
 \frac{\delta^L \Gamma}{\delta K_a(x)} 
  \; + \;
  \sum_{r\geq 2}\frac 1{r!}\,
\frac{\partial^R \Gamma}{\partial \xi^B} \, f_{A_1\cdots A_r}^{B}
\xi^{A_r}\cdots\xi^{A_1} = 0 \;.
\label{WI3}
\end{equation}

The  Ward  identities (\ref{WI1}, \ref{WI3})
capture the consequences of both the local
and the global symmetries for the generating functionals.
They hold even when the gauge fixing fermion is not invariant under 
the rigid symmetries, provided
there are no anomalies (see below).  Indeed, no condition was ever
assumed on the gauge fixing fermion, except that it should
fix the gauge.  This is of course, as it should, since the (BRST-invariant)
physical subspace yields a true representation of the symmetry \cite{npb}.

The identities on the Green functions are obtained in the usual manner,
by differentiating
(\ref{WI1}, \ref{WI3})
with respect to the sources and setting these sources equal to zero
afterwards.  In particular, since the rigid symmetries of higher
order have their own rigid ghosts, they lead to independent
identities.

%=========================================================================
\section{Example}
%=========================================================================

In order to illustrate the above formulas, consider the simple case 
of a $2$-form abelian gauge field $B_{\mu \nu}$ with classical action
\begin{equation}
S_0 = \int \mbox{d} x \;
 \left( - \frac{1}{12} F_{\mu \nu \rho} F^{\mu \nu \rho}  \right),
\quad
F_{\rho \mu \nu} = \partial_{[\rho} B_{\mu \nu]}\ .
\end{equation}
This theory is invariant not only under the well known 
first stage reducible gauge 
transformations, but also (among other rigid symmetries) 
under the following two global transformations:
\begin{equation}
B_{\mu \nu} \longrightarrow B_{\mu \nu} + 
a^{\sigma} \partial_{\sigma} B_{\mu \nu}\; ,\quad
a^\mu=constant
\label{globaltrafo1}
\end{equation}
and 
\begin{equation}
B_{\mu \nu} \longrightarrow B_{\mu \nu} + b_{\mu \nu \sigma} x^{\sigma} \; ,
\quad b_{\mu \nu \sigma}=b_{[\mu \nu \sigma]}=constant.
\label{globaltrafo2}
\end{equation}
This exactly parallels the abelian 1-form case treated in \cite{global}.

The solution of the master-equation incorporating only the gauge-symmetries
reads
\begin{equation}
S  =  \int \mbox{d} x \; \left\{ 
  - \frac{1}{12} F_{\mu \nu \rho} F^{\mu \nu \rho} + 
  B^{* \mu \nu} \left( \partial_{\mu} C_{\nu} - 
  \partial_{\nu} C_{\mu} \right) + C^{* \mu} \partial_{\mu} C \right\} \;,
\end{equation}
where the $C_{\mu}$ are the ghosts corresponding to the gauge-symmetries
and $C$ stands for the second-order ghost related to the reducibilty
of the gauge-symmetries.
In order to find the generators of the two global symmetries, one just has
to calculate the BRST invariant extension of the corresponding 
ghost-independent pieces 
$S_\sigma^0 = \int \mbox{d} x \; 
B^{* \mu \nu} \partial_{\sigma} B_{\mu \nu}$
and
$S^{0 \ \mu \nu \sigma} = \int \mbox{d} x \; 
B^{* [\mu \nu} x^{\sigma]}$.
The result is
\begin{equation}
S_{\sigma} = 
\int \mbox{d} x \; \left\{ B^{* \mu \nu} \partial_{\sigma} B_{\mu \nu} 
 + C^{* \mu} \partial_{\sigma} C_{\mu} +  
C^* \partial_{\sigma} C \right\}
\end{equation}
and
\begin{equation}
S^{\mu \nu \sigma} = 
\int \mbox{d} x \; B^{*\, [\mu \nu} x^{\sigma]} \;.
\end{equation}
In addition to the global symmetries discussed above, the model possesses
a conservation law of order three which reads explicitly 
$\partial_{\rho} F^{\rho \mu \nu} \approx 0$ and corresponds to the 
symmetry of the master-equation generated by 
\begin{equation}
S_{C^*} = \int \mbox{d} x \; C^*.
\end{equation} 
In fact, $S_{C^*}$ can be regarded as the 
representative of the non-trivial BRST-cohomology at ghost-number $-3$.

$S_{C^*}$ has to be taken into account because otherwise
the algebra generated by $S_{\sigma}$ and 
$S^{\mu \nu \sigma}$ would not close. Indeed one gets the
following antibrackets:
\begin{eqnarray}
(S_\sigma,S^{\mu\nu\rho})&=&(S,S^{\mu\nu\rho}_\sigma),
\nonumber\\
(S_{[\sigma},S^{\mu\nu\rho}_{\lambda]})
&=&(S,S^{\mu\nu\rho}_{\lambda\sigma}),
\nonumber\\
(S_{[\sigma},S^{\mu\nu\rho}_{\lambda\tau]})&=&
\delta_{[\sigma}^{\mu}\delta_\lambda^\nu\delta_{\tau]}^{\rho} S_{C^*}
\label{algebra}\end{eqnarray}
where
\begin{eqnarray*}
S^{\mu\nu\rho}_\sigma&=&
- \frac{1}{2} \int \mbox{d} x \; C^{*\, [\mu}x^\nu\delta^{\rho]}_\sigma\ ,
\\
S^{\mu\nu\rho}_{\sigma\lambda}&=&
\frac{1}{2} \int \mbox{d} x \; 
C^*x^{[\mu}\delta^\nu_\sigma\delta^{\rho]}_\lambda\ .
\end{eqnarray*}
All the antibrackets of $S^{\mu\nu\rho}$, $S^{\mu\nu\rho}_\sigma$, 
$S^{\mu\nu\rho}_{\sigma\lambda}$ and $S_{C^*}$ vanish
because these generators involve only the antifields,
but not the fields.
Together with $(S_\sigma,S_{C^*})=0$ we therefore get a
closed algebra of the type described in section \ref{structure}
with
\begin{eqnarray}
& &\{S_{\alpha_1}\}\equiv\{S_\sigma,\, S^{\mu\nu\rho},\,
S_{C^*}\}, \nonumber \\
& &\{S_{\alpha_1\alpha_2}\}\equiv\{S^{\mu\nu\rho}_\sigma\},
\quad
\{S_{\alpha_1\alpha_2\alpha_3}\}\equiv
\{S^{\mu\nu\rho}_{\sigma\lambda}\}
\end{eqnarray}
and nonvanishing structure constants 
$f_{\alpha_1\alpha_2\alpha_3\alpha_4}^\beta$ of fourth order.
Note that it is 
the last identity in (\ref{algebra}) which makes it necessary
to include $S_{C^*}$.

Introducing global ghosts $\xi^{\sigma}$, $\xi_{\mu \nu \rho}$ and $\xi$
corresponding to $S_\sigma$, $S^{\mu\nu\rho}$ and $S_{C^*}$
respectively, the extended
master-equation (\ref{extendedmaster1}) has the following form:
\begin{equation}
\left( {\cal S} , {\cal S} \right) + 
\int \mbox{d} x \; C^* 
\xi_{\mu \rho \sigma} \xi^{\mu} \xi^{\rho} \xi^{\sigma} 
= 0 \;.
\end{equation}
All the ghosts have odd Grassmann-parity. 
$\xi^{\sigma}$ and $\xi_{\mu \nu \sigma}$ have ghost-number $1$ while
$\xi$ has ghost-number $3$. The solution ${\cal S}$ reads
\begin{eqnarray}
{\cal S}  & =  & \int \mbox{d} x \; \{ 
  -\mbox{\small{$\frac{1}{12}$}} F_{\mu \nu \rho} F^{\mu \nu \rho} + 
  B^{* \mu \nu} \left( \partial_{\mu} C_{\nu} - 
  \partial_{\nu} C_{\mu} \right) + C^{* \mu} \partial_{\mu} C 
\nonumber \\
 & & + B^{* \mu \nu} 
    \left( \partial_{\sigma} B_{\mu \nu} \xi^{\sigma} + 
     x^{\sigma} \xi_{\mu \nu \sigma} \right) +
  C^{* \mu}  \partial_{\sigma} C_{\mu}\, \xi^{\sigma} + 
  C^* \left( \partial_{\sigma} C\, \xi^{\sigma} + \xi \right) 
\nonumber \\
 & & - \mbox{\small{$\frac{1}{2}$}} C^{* \mu}  
  x^{\nu} \xi_{\mu \nu \sigma} \xi^{\sigma} 
  + \mbox{\small{$\frac{1}{2}$}} C^*  
  x^{\mu} \xi_{\mu \rho \sigma} \xi^{\rho} \xi^{\sigma} \}
\end{eqnarray}
The master-equation can be cast in the standard form (\ref{extendedmaster2})
by adding to ${\cal S}$ a term depending on the global antifield $\xi^*$:
\begin{equation}
{\cal S}' = {\cal S} + 
\mbox{\small{$\frac{1}{2}$}}
\xi^* \xi_{\mu \rho \sigma} \xi^{\mu} \xi^{\rho} \xi^{\sigma} \;.
\end{equation}

The Ward identities associated with the 
higher order conservation laws may yield
useful information on the Green functions.  
For instance, in the more general case of an interacting $2$-form
abelian gauge field $B_{\mu \nu}$ with Lagrangian 
\begin{equation}
{\cal L} =
{\cal L} \left( F_{\lambda \mu \nu}, \partial_\rho F_{\lambda \mu \nu} 
\dots \right)  \;,
\end{equation} 
the conservation law of order three 
$\partial_\alpha (\delta {\cal L} / \delta F_{\alpha \mu \nu})
\approx 0$ survives. 
Also the global invariance Eq.(\ref{globaltrafo1}) remains valid,
in contrast to the second global symmetry (\ref{globaltrafo2}) which
is not a symmetry of ${\cal L}$ in general.
Dropping the latter symmetry, the analogous extended solution reads
\begin{eqnarray}        
{\cal S} & = & \int \mbox{d} x \; 
\{ {\cal L} +
B^{* \mu \nu}
\left( \partial_{\mu} C_{\nu} - \partial_{\nu} C_{\mu} \right) + 
C^{* \mu} \partial_{\mu} C  \nonumber \\
& & +  B^{* \mu \nu}  \partial_{\sigma} B_{\mu \nu} \xi^{\sigma} + 
C^{* \mu} \partial_{\sigma} C_{\mu} \xi^{\sigma} + 
C^* \left( \partial_{\sigma} C \xi^{\sigma} + \xi \right) \}  
\end{eqnarray} 
and the Ward identity (\ref{WI1}) then implies
\begin{equation}
\int \mbox{d} x \; J_{(C)}(x) \left< \partial_{\sigma} C(x) \xi^{\sigma} + 
\xi \right> = 0 \;.
\end{equation}
The ($\xi$-dependent part of) identity (\ref{WI3})
for the effective action implies
\begin{equation}
\int \mbox{d} x \; \frac{\delta \Gamma }{\delta C(x)} = 0,
\label{example}
\end{equation}
where $\Gamma = \Gamma(K=0, \xi =0)$.  
The identity
(\ref{example}) expresses the invariance of the effective action
$\Gamma$ under constant
shifts of the ghost of ghost $C(x)$.

%=========================================================================
\section{Anomalies}
%=========================================================================

Let us finally discuss some applications of the resulting extended
antifield formalism that parallel analogous applications
of the usual BRST formalism.
Consider for instance the problem of anomalies in the
rigid symmetries. They can be analysed along the algebraic 
lines initiated in the pioneering work \cite{BRS}.
The procedure is explained in \cite{Bonneau,Blasi}, and we just recall here the
main arguments, taking into account the higher order symmetries.
An anomaly appears as a
violation of the master equation (\ref{WI3}) for
the regularized $\Gamma$
and must fulfill, to lowest loop order,
the generalized Wess-Zumino consistency condition
\cite{wz}
\begin{equation}
D \int \mbox{d} x \, a = 0
\label{consistency}
\end{equation}
where the extended BRST differential $D$ is defined
(in the functional space ${\cal E}$) by
\begin{equation}
D X \equiv (X,{\cal S}')'\quad\Rightarrow\quad D^2=0.
\label{2A}
\end{equation}
and takes into account {\it all} symmetries.
For local functionals not
involving the $\xi^*_A$, like $\int \mbox{d} x \, a$, $D$ takes
the form
\begin{equation}
X=\int \mbox{d} x \, a\ \Rightarrow\ 
D X = (X,{\cal S}) + \sum_{r\geq 2}\frac 1{r!}\,
\frac{\partial^R X}{\partial \xi^B} \, f_{A_1\cdots A_r}^{B}
\xi^{A_r}\cdots\xi^{A_1} \;.
\label{2}
\end{equation}
An anomaly (for the local or global symmetries, or combinations thereof)
is a solution of (\ref{consistency}) that is non-trivial, i.e. not of
the exact form $D\int \mbox{d} x \, b$.  The investigation of the possible
anomalies in the local and global symmetries of all orders is
accordingly equivalent to the problem of computing the
cohomology of the extended BRST differential $D$ at ghost number one.
Terms proportional to the rigid ghosts would define anomalies in the
corresponding rigid symmetries.

The problem of computing the cohomology of $D$ in the space of local
functionals is equivalent to the problem of computing the cohomology
of $D$ modulo the spacetime exterior derivative in the space
of local volume forms, provided the fields decrease fast
enough at infinity. (This excludes instanton-like configurations,
\cite{wein,schwarz,hooft}; see also \cite{braga}.)

%=========================================================================
\section{Deformations}
%=========================================================================

Consider next the question of
whether it is possible to deform a given action
continuously such that it remains invariant under (possibly deformed)
gauge and global symmetries. This problem can be efficiently
and systematically studied in the antifield formalism along
the lines of \cite{bh,stasheff} by looking for a solution to the
extended master equation of the form
\begin{equation} 
{\cal S}'_\tau={\cal S}'+ \tau {\cal S}^{(1)}{}'+
\tau^2 {\cal S}^{(2)}{}'+\ldots
\label{20a}
\end{equation}
where ${\cal S}'$ is the original (undeformed) solution to the
extended master equation containing the gauge
symmetries and the global symmetries in question,
and $\tau$ is  a constant deformation parameter. To first
order in $\tau$ the extended master equation for ${\cal S}'_\tau$
then requires
\begin{equation} 
\left({\cal S}^{(1)}{}',{\cal S}'\right)'\equiv
D\, {\cal S}^{(1)}{}' =0.
\label{21a}
\end{equation}
The
first order deformation ${\cal S}^{(1)}{}'$ is thus invariant under
the extended BRST operator $D$.
This allows to classify the possible deformations in question
(to first order in $\tau$)
through investigating the $D$-cohomology at ghost number 0.
Note that we have used the extended master equation
in its compact form (\ref{extendedmaster2}).
This is particularly useful in this context because it
automatically takes into account that some of the structure
constants $f_{A_1\cdots A_r}^B$ may get deformed too.

In general it will be
neither possible nor desirable to promote {\em all} the global
symmetries of the original action to symmetries of a nontrivially
deformed action. Therefore in general
one would actually include only a physically important
subset of the global symmetries
in ${\cal S}$. An instructive example is the deformation of
free abelian gauge theories to Yang--Mills theories which
was discussed in \cite{bht} along the lines of \cite{bh}. Clearly
Yang--Mills theories have less global symmetries than the
corresponding free abelian gauge theories (which have in fact
infinitely many nontrivial global symmetries), showing that
indeed not all the global symmetries can be maintained.
However, it is evidently possible
to keep at least the physically important Poincar\'e symmetries.

A similar application is the construction and
classification of actions which
are invariant under prescribed gauge and global symmetries
whose commutator algebra closes off-shell.
This problem can be studied through the $D$-cohomology
at ghost number 0 too. An example for such an
investigation (rigid N=1 supersymmetry in four dimensions)
can be found in \cite{glusy}.

%=========================================================================
\section{Conclusions}
%=========================================================================

We have developed in this paper the master equation formalism
for both local and rigid symmetries.  The key to overcoming
difficulties encountered in the past is to introduce global
ghosts for all the rigid symmetries, and not just for those of first order.
We have shown explicitly how to incorporate in an appropriately
extended master equation the
higher order rigid symmetries associated with higher order conservation
laws $\partial_{\mu_1}j^{\mu_1\cdots\mu_k}\approx 0$.
This leads to new Ward identities and, more importantly, avoids
the obstructions that may be encountered when trying to construct a 
solution of the extended master equation (\ref{extended'}) that does not
take these higher order conservation laws into account.  While
(\ref{extendedmaster1}) is never obstructed,
(\ref{extended'}) may fail to have local solutions in the presence of higher
order symmetries.  Of course, (\ref{extended'})
will not be obstructed if a subset of first order conservation
laws is used that defines a subalgebra in the above sense. 
But otherwise  obstructions can -- and do -- arise
\cite{global}.  Furthermore,  the
Ward identities associated with the 
higher order conservation laws may yield
useful information on the Green functions.
Finally, the extended antifield approach turns out to be 
useful in the systematic analysis by cohomological techniques
of deformation problems or anomaly issues.

{\em Acknowledgements:}
This work has been supported  in part by research
funds from the F.N.R.S.
(Belgium) and research contracts with the Commission of the European
Community. F.B. has been supported by the Spanish
ministry of education and science (MEC).

\end{document}